\documentclass[12pt]{article}
\usepackage{authblk}
\usepackage[bookmarksnumbered, colorlinks, plainpages]{hyperref}
\usepackage{amsmath, amsthm, amscd, amsfonts, amssymb, graphicx, color, booktabs,cite,url}
\newcommand{\eps}{\epsilon}

\newcommand{\mpi}{M_\pi}
\newcommand{\mpic}{M_{\pi^\pm}}

\newcommand{\beq}{\begin{equation}}
\newcommand{\eeq}{\end{equation}}

\providecommand{\MeV}{\,\text{MeV}}

\numberwithin{equation}{section}
\definecolor{email}{rgb}{0.00,0.00,0.84}
\begin{document}
\setcounter{page}{1}

\title{\large \bf 12th Workshop on the CKM Unitarity Triangle\\ Santiago de Compostela, 18-22 September 2023 \\ \vspace{0.3cm}
\LARGE Prospects for PIONEER}

\author[1]{Martin Hoferichter\footnote{hoferichter@itp.unibe.ch} on behalf of the PIONEER collaboration}
\affil[1]{Albert Einstein Center for Fundamental Physics,\linebreak Institute for Theoretical Physics, University of Bern,\linebreak Sidlerstrasse 5, 3012 Bern, Switzerland}
\date{}

\maketitle

\begin{abstract}
Pion $\beta$ decay, $\pi^+ \to\pi^0 e^+ \nu_e$, offers a pristine way to measure the CKM matrix element $V_{ud}$ in a purely mesonic system, with excellent control over the hadronic matrix elements. We review the physics goals and current status of the PIONEER experiment, which aims at major improvements in the branching fractions for the $\pi^+\to e^+ \nu_e$ decay in Phase I and for pion $\beta$ decay in Phases II and III of its experimental program, potentially leading to a  measurement of $V_{ud}$ competitive with determinations from $\beta$ decays involving nucleons.  
\end{abstract} 

\maketitle

\section{Introduction}

PIONEER is a next-generation rare pion decay experiment at PSI~\cite{PIONEER:2022yag,PIONEER:2022alm}. Its main physics goals include the test of lepton flavor universality in the ratio
\beq
   R_{e/\mu}=\frac{\Gamma[\pi^+\to e^+\nu_e(\gamma)]}{\Gamma[\pi^+\to \mu^+\nu_\mu(\gamma)]}
   \eeq
at the level of $10^{-4}$ (Phase I), the test of CKM unitarity by measuring  
$V_{ud}$ at $3\times 10^{-4}$ in pion $\beta$ decay (Phases II+III), and searches for exotic new particles, e.g., sterile neutrinos. The experiment has been approved at PSI and is currently in its R\&D stage~\cite{PIONEER:2023ida,Iwamoto:2023uov,Mazza:2021adt}. 

\section{Physics goals}
\label{sec:physics}

\subsection{Lepton flavor universality}

$R_{e/\mu}$ is arguably the most precisely predicted  observable involving hadrons~\cite{Marciano:1993sh,Cirigliano:2007xi,Cirigliano:2007ga}
\beq
\label{RSM}
R_{e/\mu}^\text{SM}= 1.23524(15)\times 10^{-4}, 
\eeq
with a remaining uncertainty that quantifies the impact of unresolved hadronic effects, parameterized via the low-energy constants that arise at two-loop level in chiral perturbation theory. The experimental value~\cite{ParticleDataGroup:2022pth,PIENU:2015seu,Czapek:1993kc,Britton:1992pg,Bryman:1985bv}
\beq
   R_{e/\mu}^\text{exp}=1.2327(23)\times 10^{-4}
   \eeq
is dominated by the PIENU experiment~\cite{PIENU:2015seu}, and while small improvements are expected from PEN~\cite{PEN:2018kgj} and the full PIENU data set~\cite{PIENU:2018lsf}, this will not suffice to challenge the Standard-Model prediction~\eqref{RSM}. However, the experience gained with the previous experiments critically informs the PIONEER design. 

Due to a chiral enhancement with $\mpi^2/[m_e(m_u+m_d)]$, $R_{e/\mu}$ is particularly sensitive to (pseudo-)scalar currents, so that a test at the level of $10^{-4}$ can probe scales up to several PeV~\cite{Bryman:2011zz}. Moreover, already now $R_{e/\mu}$ gives one of the best constraints on modified $W$ couplings~\cite{Crivellin:2020lzu,Bryman:2021teu}
\beq
   \frac{R_{e/\mu}^\text{SM}}{R_{e/\mu}^\text{exp}}=1+\eps_{\mu\mu}-\eps_{ee}=1.0010(9),
   \eeq
opening another order of magnitude in parameter space once the experimental precision reaches Eq.~\eqref{RSM}. This is particularly intriguing in view of other hints for the violation of lepton flavor universality~\cite{Crivellin:2021sff}, including a possible connection to a deficit in the unitarity relation for the first row of the CKM matrix~\cite{Crivellin:2020lzu,Crivellin:2021njn}.

\subsection{CKM unitarity}

\begin{figure}[t]
	\centering
	\includegraphics[width=0.5\linewidth,clip]{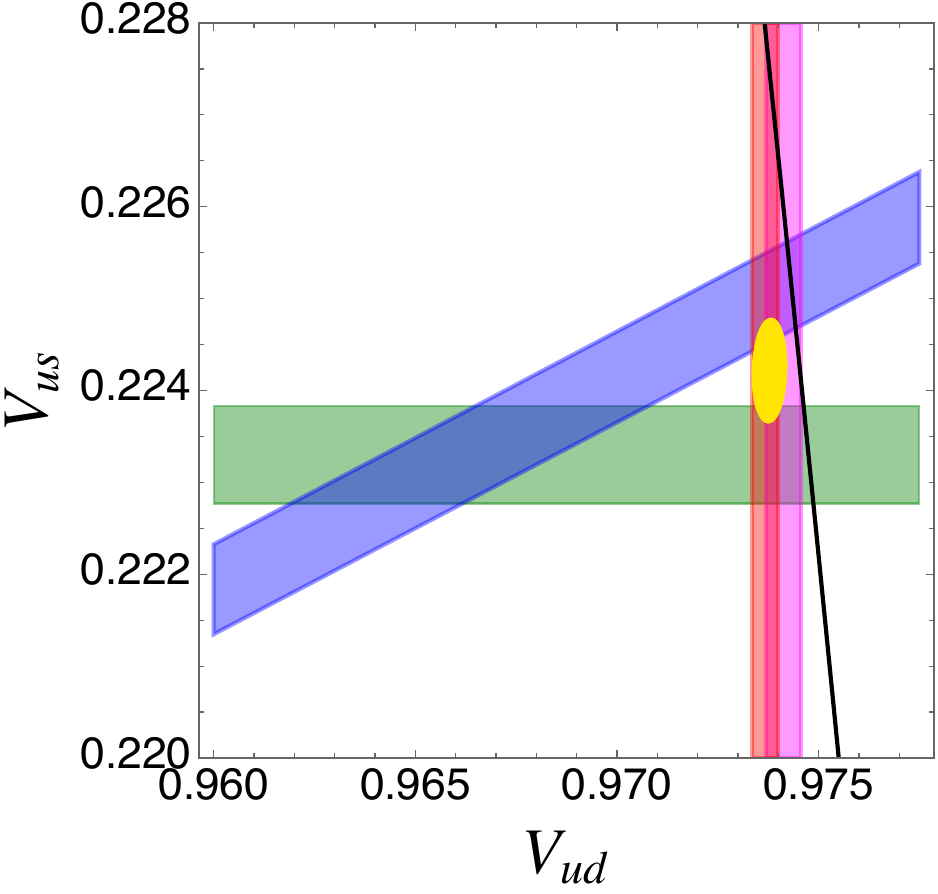}
	\caption{Current status of first-row CKM unitarity, figure taken from Ref.~\cite{Cirigliano:2022yyo}.
	The vertical bands give constraints from superallowed $\beta$ decays (left, red) and neutron decay (right, violet), 
	the horizontal band from $K_{\ell 3}$ decays (green), and the diagonal band from $K_{\ell 2}/\pi_{\ell 2}$ decays (blue). Unitarity is represented by the black line and the global fit by the yellow ellipse, whose tension evaluates to $2.8\sigma$.}
	\label{fig:CKM}
\end{figure}

Besides the possible relation to lepton flavor universality, the most immediate impact PIONEER will have on the CKM matrix concerns Phases II+III via an improved measurement of pion $\beta$ decay. The current status of the first-row unitarity test 
\beq
|V_{ud}|^2+|V_{us}|^2+|V_{ub}|^2=1
\eeq
is shown in Fig.~\ref{fig:CKM}, indicating a unitarity tension for the intersection of the $K_{\ell 2}$ and $K_{\ell 3}$ bands of $2.6\sigma$, and $2.8\sigma$ for the global fit~\cite{Cirigliano:2022yyo}. While the tension involving $V_{us}$ calls for a new measurement of $K_{\mu 3}/K_{\mu2}$ decays~\cite{Cirigliano:2022yyo}, several avenues exist to obtain improved determinations of $V_{ud}$. 

The nominally best value is obtained from superallowed $\beta$ decays~\cite{Hardy:2020qwl}, in which case the uncertainties are dominated by nuclear corrections~\cite{Gorchtein:2018fxl,Gorchtein:2023naa}, but improved theory is being developed, see, e.g., Ref.~\cite{Cirigliano:2023fnz}. The precision of $V_{ud}$ from neutron decay is getting close if the current best measurements of the lifetime~\cite{UCNt:2021pcg} and the asymmetry $\lambda=g_A/g_V$~\cite{Markisch:2018ndu} are taken, but the
interpretation is complicated by a scale factor to account for the tension with Ref.~\cite{Beck:2019xye}. As a third possible determination, pion $\beta$ decay would allow one to obtain $V_{ud}$ from a purely mesonic system, via the master formula~\cite{Cirigliano:2002ng,Czarnecki:2019iwz} 
 \beq
   \Gamma[\pi^+\to\pi^0 e^+\nu_e(\gamma)]=
\frac{G_F^2|V_{ud}|^2\mpic^5|f_+^\pi(0)|^2}{64\pi^3}(1+\Delta_\text{RC}^{\pi\ell})I_{\pi\ell}, 
\eeq
where $G_F$ is the Fermi constant from muon decay~\cite{MuLan:2012sih},  $I_{\pi\ell}=7.3767(41)\times 10^{-8}$ is the phase-space factor, $\Delta_\text{RC}^{\pi\ell}$ denotes radiative corrections, and $f_+^\pi(0)=1-7\times10^{-6}$~\cite{Cirigliano:2002ng} the $SU(2)$ Ademollo--Gatto-protected~\cite{Ademollo:1964sr,Behrends:1960nf} form-factor normalization. Using the radiative corrections $\Delta_\text{RC}^{\pi\ell}=0.0332(3)$~\cite{Feng:2020zdc}, the current measurement from PIBETA~\cite{Pocanic:2003pf} yields
\beq
    V_{ud}=0.97386(281)_\text{BR}(9)_{\tau_\pi}(14)_{\Delta_\text{RC}^{\pi\ell}}(28)_{I_{\pi\ell}}[283]_\text{total},
   \eeq
where the branching ratio (BR) by far dominates the uncertainty, presenting another opportunity for an order-of-magnitude improvement before theory uncertainties become relevant. In fact, the second-largest uncertainty from the phase-space factor $I_{\pi\ell}$ derives from the present knowledge of the pion mass difference $M_{\pi^\pm}-M_{\pi^0}=4.59364(48)\MeV$~\cite{Crawford:1990jc}, which could potentially be improved with a dedicated experiment. The goal for Phase II of PIONEER is to improve the BR by a factor $3$, obtaining a competitive measurement of $V_{ud}/V_{us}$ from $\pi_{\ell3}/K_{\ell 3}$ decays, while aiming for the ultimate precision of $3\times 10^{-4}$ in Phase III.

\subsection{Exotics}

\begin{figure}[t]
	\centering
	\includegraphics[width=0.9\linewidth,clip]{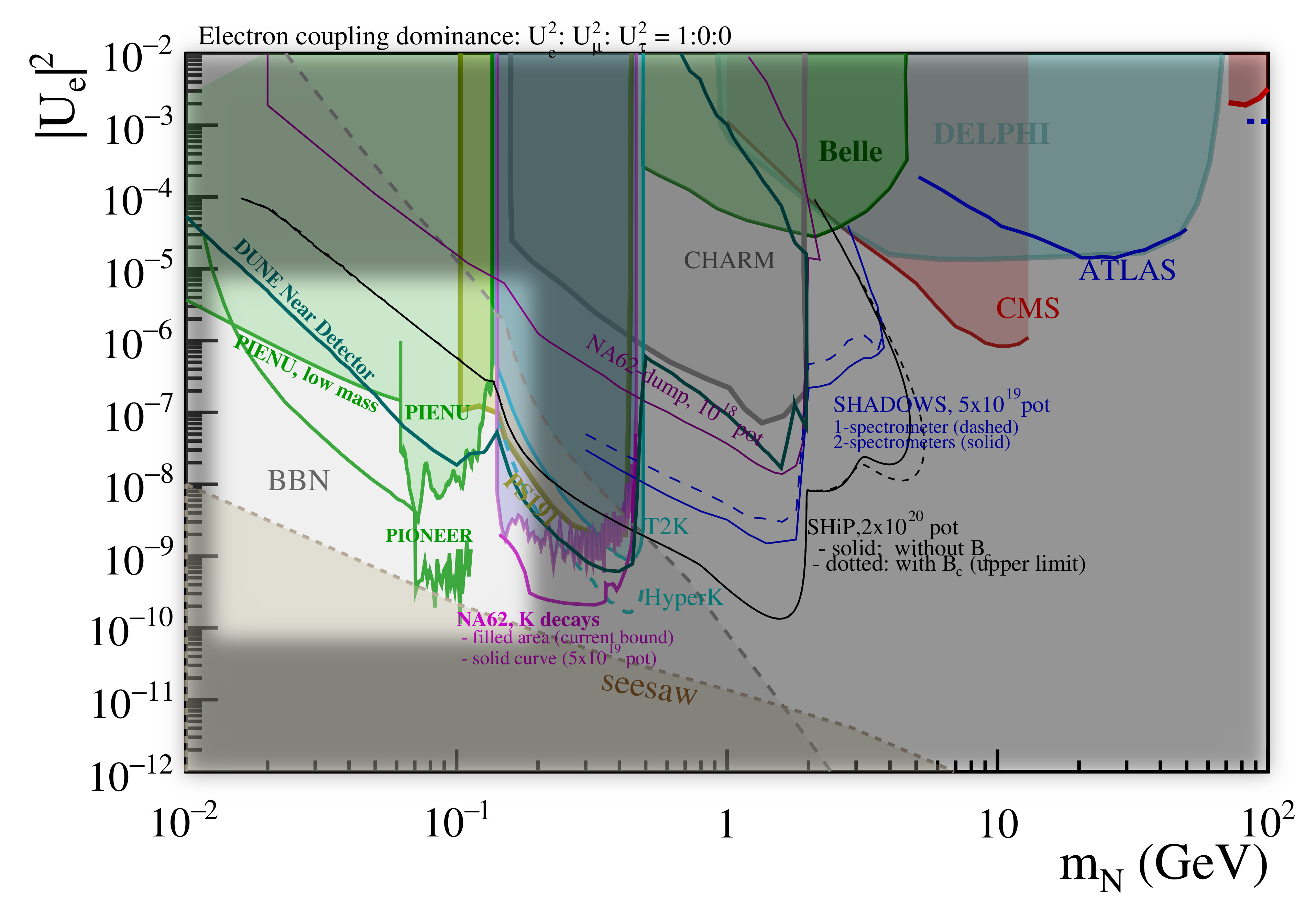}
	\caption{Exclusion plot for coupling strength vs.\ mass of a new heavy neutrino, figure taken from Ref.~\cite{Abdullahi:2022jlv}.}
	\label{fig:neutrino}
\end{figure}

Finally, PIONEER is poised to make major contributions in the search for heavy neutrinos $\nu_h$ and other dark-sector physics, see Fig.~\ref{fig:neutrino}, e.g., by peak searches in the positron energy spectrum~\cite{PIENU:2017wbj}. PIENU has also searched for $\pi^+\to\mu^+\nu_h$~\cite{PIENU:2019usb}, $\mu^+\to e^+ X$~\cite{PIENU:2020loi}, and $\pi^+\to\ell^+\nu X$~\cite{PIENU:2021clt}, and PIONEER should be able to improve the sensitivity by an order of magnitude in either channel.  

\section{Current status}

PIONEER builds upon the legacy of the PIENU, PEN, and PIBETA experiments~\cite{Frlez:2003vg,PIENU:2015pkq},
as the experience gained therein indicates the improvements needed to achieve the ambitious goals outlined in Sec.~\ref{sec:physics}. Crucially, these include (i) a segmented active target (ATAR) allowing for 5D tracking (energy, time, and three spatial dimensions), built from silicon-strip, low-gain avalanche detectors (LGADs); (ii) a $3\pi$, $25$ radiation-length electromagnetic calorimeter (Calo) with energy resolution $\delta E/E\leq 1.5\%$, either using liquid xenon or LYSO crystal technology.  

\subsection{Experimental principle}

\begin{figure}[t]
	\centering
	\includegraphics[width=0.54\linewidth,clip]{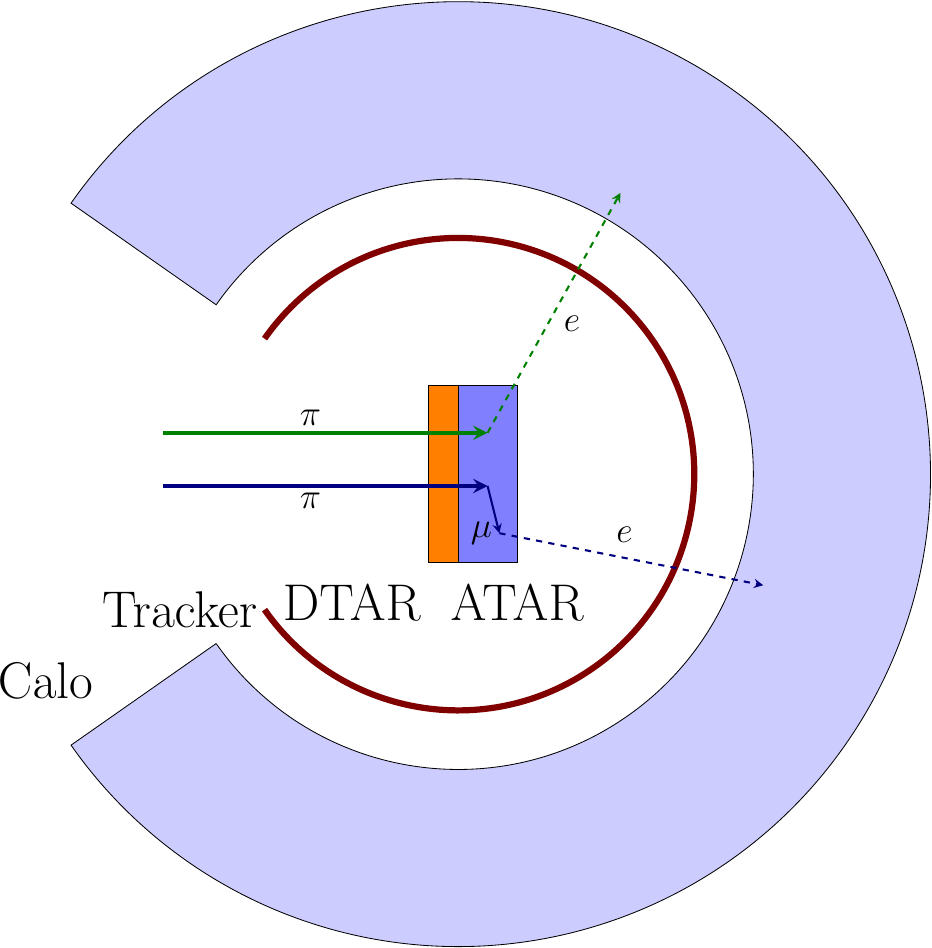}
	\caption{Sketch of the PIONEER experiment. The pion beam is first slowed down in the degrader target (DTAR) before reaching the ATAR. }
	\label{fig:principle}
\end{figure}

The basic principle of the experiment is depicted in Fig.~\ref{fig:principle}, the key idea being to ``count and sort'' the positrons emitted by the stopped pions, either from the direct decay $\pi^+\to e^+\nu_e$ or the muonic mode with subsequent Michel decay $\pi^+\to\mu^+\nu_\mu\to e^+\nu_e\nu_\mu\bar\nu_\mu$. In the ratio $R_{e/\mu}$ many systematic effects will cancel.

\begin{figure}[t]
	\centering
	\includegraphics[width=\linewidth,clip]{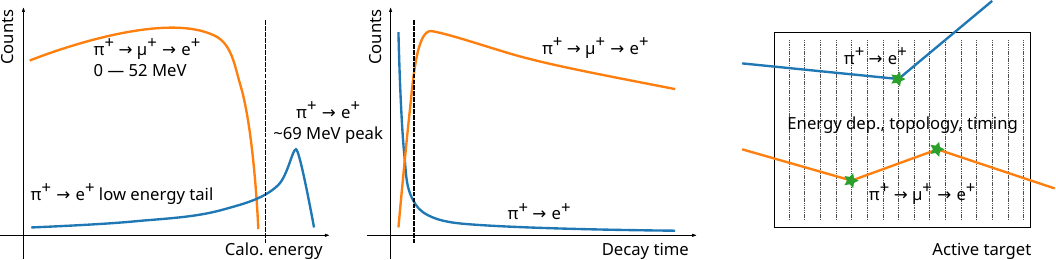}
	\caption{Signatures of the two decay modes sketched in Fig.~\ref{fig:principle} in the calorimeter (left), the ATAR (right), and 
	their decay-time spectra (middle). }
	\label{fig:spectra}
\end{figure}

To distinguish the two processes, a combination of the different detector components is used, see Fig.~\ref{fig:spectra}. First, the signature in the calorimeter isolates the main $\pi^+\to e^+$ peak  at $\simeq 69\MeV$, but at the intended level of precision a detailed understanding of the low-energy tail is key, which in turn mandates a sufficient number of radiation lengths $X_0\geq  25$ and energy resolution $\delta E/E\leq 1.5\%$. Both strategies using liquid xenon and LYSO crystal technology are 
currently under investigation to achieve these specifics. 

Second, the two decay modes can be distinguished in the ATAR using energy deposition, event topology, and timing information, for the latter the spectra are also included in Fig.~\ref{fig:spectra}. This detailed tracking information is an essential improvement over the PEN and PIENU setups. 

\subsection{Error budget}

Based on the improvements in the proposed detection system described above, PIONEER aims for an overall uncertainty ${\mathcal O}(0.01\%)$. This estimate is based on $2 \times 10^8$ $\pi^+ \to e^+ \nu_e$ events, corresponding to three 5-month runs. Major improvements are projected across all sources of error, including improved statistics and a reduced low-energy $e^+$ tail. 

\subsection{Time line}

PIONEER was approved at PSI in 2022~\cite{PIONEER:2022yag}, and is currently in its R\&D stage.  
The first physics runs could take place in the early 2030s.  

\section{Conclusions}

PIONEER is an ambitious next-generation experiment to measure rare pion decays, aiming for a $10^{-4}$ test of $R_{e/\mu}$ in Phase I, a competitive determination of $V_{ud}$ from pion $\beta$ decay in Phases II+III, and improved sensitivity to exotic decays. In both phases, the experiment targets processes in which the theoretical prediction currently surpasses experiment by an order of magnitude, presenting an opportunity for a clean discovery. To this end, the PIONEER collaboration employs state-of-the-art technologies, with members from a diverse group of experiments including PIENU, PEN, NA62, MEG, E989, and HEP colliders. 


\section*{Acknowledgments}
 Financial support by the SNSF (Project No.\ PCEFP2\_181117) is gratefully acknowledged.

\bibliographystyle{JHEP}
\bibliography{References}

\end{document}